\documentclass[english,aps,prl,a4paper,twocolumn]{revtex4}
\usepackage[T1]{fontenc}
\usepackage{geometry}
\usepackage{mathrsfs}
\usepackage{color}
\usepackage{amsmath}
\usepackage{amsthm}
\usepackage{subfig}
\usepackage{graphicx,epsf}
\usepackage{epstopdf}
\usepackage{babel}



\usepackage{babel}
\begin{document}
\title{Drift wave solitons and zonal flows: implication on staircase formation
}
\author{Ningfei Chen$^{1,2}$, Liu Chen$^{2,3}$, Fulvio Zonca$^{2,3}$ and Zhiyong Qiu$^{1,3}$\footnote{Author to whom correspondence should be addressed: zqiu@ipp.ac.cn}}

\affiliation{$^1$Key Laboratory of Frontier Physics in Controlled Nuclear Fusion and Institute of Plasma Physics, Chinese Academy of Sciences,
Hefei 230031, China\\
$^2$Institute for  Fusion Theory and Simulation, School of Physics, Zhejiang University, Hangzhou 310027, China\\
$^3$Center for Nonlinear Plasma Science and   C.R. ENEA Frascati, C.P. 65, 00044 Frascati, Italy
}

\begin{abstract}
The self-consistent nonlinear interaction of drift wave (DW) and zonal
flow (ZF) is investigated using nonlinear gyrokinetic theory, with both spontaneous
excitation and beat-driven of ZF by DW treated on the same footing. DW solitons are formed in the nonlinear DW-ZF interactions and are confined between radially spaced micro-barriers. The resulting radial structures in the nonlinear DW-ZF interactions exhibit similar pattern to the $\mathbf{E}\times \mathbf{B}$ ``staircase'' observed in numerical simulations. These micro-barriers are generated by the repulsive response due to   spontaneously excited
ZF, which, as a general property demonstrated in this work, also generate an attractive nonlinear potential in DW equation. Meanwhile, the nonlinear potential due to beat-driven ZF is always attractive and, as such, always serve as potential well to contribute
to soliton formation. For spontaneously excited
ZF from initial noise, the simultaneous excitation of solitons and micro-barriers is found to be universal, due to the zero frequency nature of ZF and spatial   structure
of the Reynolds stress. The present analysis, thus, provides a potential first-principle-based interpretation of the $\mathbf{E}\times \mathbf{B}$ staircase observed in simulations, which may contribute to micro transport barriers formation and enhance plasma confinement.
\end{abstract}
\maketitle

Self-organizing processes \cite{PBakPRL1987,PBakPRA1988} are common nonlinear phenomena in geophysics \cite{DSornetteJGR1990}, optics \cite{AHasegawaAPL1973A}, plasma physics \cite{AHasegawaAP1985} and other fields. In the context of thermal plasma physics, it is observed that a self-organizing $\mathbf{E}\times \mathbf{B}$ flow pattern is generated in numerical simulations, which is called ``staircase'' and turns out to change equilibrium
temperature profile significantly \cite{GPradalierPRL2015,GPradalierNF2017}. Staircase is a global, long-lived
self-organizing phenomenon during which smooth  temperature profile evolves
into spatially quasi-regular step-like structure due to similarly distributed
micro-barriers, which may improve thermal plasma confinement. More specifically, in regions between neighboring
micro-barriers, avalanche-like transport can be observed, which flattens
the pressure profile to form the ``tread''; meanwhile, heat and
particle fluxes are reduced at micro-barriers, leading to the formation
of  \hspace{0.1em} ``steps''. The significant transport in treads is induced by
drift wave (DW) turbulence, which can be driven unstable by pressure
gradient intrinsic to magnetically confined plasmas \cite{WHortonRMP1999}. On the other
hand, these micro-barriers are believed to be resulted from
meso-scale zonal flow (ZF) \cite{PDiamondPPCF2005}. The formation of staircase structures manifests
that micro-scopic DW can be regulated by meso-scale ZF to significantly
affect  global pressure profile. Despite its crucial role in understanding self-organizing
processes in plasmas and determining overall plasma performance, the
mechanism for staircase formation is still under active investigation \cite{XGarbetPoP2021,LQiNF2022}.

Another type of self-organizing process induced by DW-ZF interactions is soliton formation \cite{ZGuoPRL2009,NChenPoP2024}, which is shown to  occur as the
dispersiveness of DW balances  nonlinear
trapping effect induced by self-generated  ZF. Soliton is characterized by  preserving   its amplitude
and shape unchanged during propagation. Thus, it may contribute to
nonlocal transport via turbulence spreading,  leading   to breakdown
of local transport models \cite{XGarbetNF1994,FZoncaPoP2004,TSHahmPPCF2004,TSHahmPoP2005,DPradalierPRE2010}. 
This is also the case for dispersionless propagation of  optical soliton in optical fiber to achieve lossless optical communication \cite{AHasegawaAPL1973A,AHasegawaAPL1973B}.
In recent works, we have demonstrated that DW solitons can form in the DW-ZF system taking into account two mechanisms for ZF excitation by DW in the fully nonlinear stage \cite{ZGuoPRL2009,NChenPoP2024}. The first one is the ZF spontaneous excitation   via modulational instability \cite{LChenPoP2000}, 
which requires DW intensity to exceed a threshold and also incorporates scattering of DW into short wavelength stable
domain, leading to self-regulation and saturation of DW turbulence.
The other one is the thresholdless ZF beat-driven excitation \cite{ZQiuPoP2016}, with the
growth rate of ZF being twice the instantaneous growth rate
of DW \cite{YTodoNF2010,GDongPoP2019}. Recently, a theoretical analysis of the DW-ZF system with both ZF generation mechanisms
accounted for on the same footing was developed, showing significant reduction
of the threshold for modulational instability due to beat-driven
ZF \cite{LChenPoP2024}. This suggests that soliton formation  may also be significantly modified if both mechanisms for ZF generation are accounted for. 
Consequently, for both academic interest and its implications to the confinement in fusion reactors, it is important to investigate soliton dynamics in the fully nonlinear DW-ZF system, accounting for both routes to ZF generation by DW.

In this work, we consider the excitation of ZF by single-$n$ electrostatic
DW in toroidal geometry, with beat-driven ZF (BZF)
and spontaneously excited ZF (SZF) being treated on the same footing. Here, $n$ is the toroidal mode number.
The two-field description of nonlinear DW-ZF system is adopted, in which DW is treated as a whole \cite{ZGuoPRL2009},
instead of seperating into pump wave and relatively small sidebands \cite{LChenPoP2000}.
Single-$n$ DW and ZF fluctuations can be expressed as
\begin{eqnarray*}
\delta\phi_{n} & = & A_{n}(r,t){\rm e}^{{\rm i}n\zeta-{\rm i}\omega_{n}t}\sum_{m}{\rm e}^{-{\rm i}m\theta}\Phi(nq-m)+c.c,\\
\delta\phi_{{\rm Z}} & = & A_{{\rm Z}}(r,t){\rm e}^{{\rm i}k_{r{\rm Z}}r-{\rm i}\omega_{{\rm Z}}t}+c.c,
\end{eqnarray*}
where $A_{n}$ and $\omega_n$ are the meso-scale radial envelope and frequency of DW, $A_{\rm Z}$ and $\omega_{\rm Z}$ are the meso-scale radial structure and frequency of ZF, which vanishes for the zero frequency component considered here \cite{PDiamondPPCF2005}, $\Phi(nq-m)$ is the micro radial
 structure due to parallel wavenumber ($k_\parallel$) spectrum, $m$
is the poloidal mode number, $q$ is the safety factor, and $c.c$ represents the complex
conjugate. The subscripts ``$n$'' and ``${\rm Z}$'' represent
the quantities associated with DW and ZF, respectively, and $k_r\equiv -{\rm i}\partial_r$ is the operator for radial derivative.  For simplicity
of disccussion, a large aspect ratio tokamak with concentric circular
magnetic surfaces is considered, and the magnetic field is given by
$\boldsymbol{B}=B_{0}[{\rm \boldsymbol{e}}_{\xi}/(1+\epsilon\cos\theta)+\epsilon{\rm \boldsymbol{e}}_{\theta}/q]$,
where $\epsilon\equiv r/R\ll1$ is the inverse aspect ratio, 
$R$ and $r$ are the major and minor radii of tokamak, while
$\xi$ and $\theta$ are the toroidal and poloidal angles, respectively. Without loss of generality, constant temperatures are considered for different particle species, such that plasma nonuniformity is accounted for by density nonuniformity. Particle responses to low-frequency
fluctuations, such as DW and ZF investigated here, are governed by
nonlinear gyrokinetic equation \cite{EFriemanPoF1982}
\begin{eqnarray}
\left(\omega+{\rm i}v_{\parallel}\partial_{l}+\omega_{{\rm Ds}}\right)\delta H_{k,{\rm s}} & = & \dfrac{q_{{\rm s}}}{T_{{\rm s}}}\left(\omega-\omega_{*{\rm s}}\right){\rm J}_{k}\delta\phi_{k}F_{0{\rm s}}\nonumber \\
 &  & -{\rm i}\dfrac{c}{B_{0}}\Lambda_{k',k''}^{k}{\rm J}_{k'}\delta\phi_{k'}\delta H_{k''{\rm ,s}}. \label{eq:NLGKE}
\end{eqnarray}
Here, $\omega_{*{\rm s}}\equiv k_{\theta}cT_{{\rm s}}/(eBL_{N})$
is the diamagnetic frequency accounting for density nonuniformity, $L_{N}\equiv-N/(\partial N/\partial r)$
is the characteristic length of density nonuniformity, $N$ is the
equilibrium particle density,  $\omega_{\rm Ds}\equiv \hat{\omega}_{{\rm ds}}C$ is the magnetic drift frequency,  with  $\hat{\omega}_{{\rm ds}}\equiv\omega_{{\rm ds}}\left(v_{\perp}^{2}/2+v_{\parallel}^{2}\right) /v^2_{\rm ts}$,
$C\equiv\cos\theta-\sin\theta k_{r}/k_{\theta}$, $\omega_{\rm ds}\equiv k_\theta cT_{\rm s}/(eBR)$,  and $v_{{\rm ts}}\equiv\sqrt{2T_{{\rm s}}/m_{{\rm s}}}$ is the thermal
velocity. Furthermore, $\delta H_{k,{\rm s}}$ is the nonadiabatic particle response, $F_{\rm 0s}$ is the equilibrium distribution function, which is taken here as Maxwellian,  ${\rm J}_k\equiv{\rm J}_{k}\left(k_{\perp}\rho_s\right)$
is the Bessel function of zero index describing finite Larmor radius (FLR) effect, $k_\perp=\sqrt{k_r^2+k_{\theta}^2}$ is the perpendicular wavenumber, $\rho_{{\rm s}}\equiv v_\perp/\omega_{{\rm cs}}$ is the Larmor radius,
 and $\omega_{{\rm cs}}\equiv eB/\left(m_{{\rm s}}c\right)$ is the cyclotron frequency.  Subscripts $k$, $k'$, $k''$ denotes
DW or ZF, and $s=i/e$ represents particle species. The second term on the right hand side of Eq. (\ref{eq:NLGKE}) is the  perpendicular
nonlinear term, with the selection rules on frequency and wavenumber matching conditions
for mode-mode coupling applied, where $\Lambda_{k',k''}^{k}\equiv\boldsymbol{b}\cdot\left(\boldsymbol{k}''\times\boldsymbol{k}'\right)$, $\boldsymbol{b}\equiv\boldsymbol{B}/B_0$ is the unit vector along equilibrium magnetic field lines, and other notations are standard.
From that term, the excitation of ZF by DW can be identified and found
to have two routes, one is the beat-driven excitation, with $k_{r}'=k_{r}''$ \cite{NChenPoP2024}
and the other is spontaneous excitation with $k_{r}'\neq k_{r}''$ \cite{LChenPoP2000}. Consequently, ZF can be written as $\delta\phi_{{\rm Z}}\equiv\delta\phi_{{\rm Z,S}}+\delta\phi_{{\rm Z,B}}$,
with two components representing the  scalar potential of SZF
and BZF, respectively. The field equations are closed
by the charge quasi-neutrality condition
\begin{eqnarray}
\dfrac{e^{2}N}{T_{{\rm i}}}\left(1+\dfrac{1}{\tau}\right)\delta\phi_{k} & = & \left\langle e{\rm J}_{0}\delta H_{{\rm i}}\right\rangle _{k}-\left\langle e\delta H_{{\rm e}}\right\rangle _{k}.\label{eq:QN}
\end{eqnarray}
Here, $\tau\equiv T_{\rm e}/T_{\rm i}$ is the ratio of electron and ion temperature. With the particle responses derived from nonlinear gyrokinetic equation
and substituted  into Eq. (\ref{eq:QN}), the set of coupled
nonlinear equations describing the nonlinear interactions between   
ZF and DW are given as
\begin{eqnarray}
 & \omega_{n}\mathscr{E}_{{\rm d}}\delta\phi_{n}=-\dfrac{c\tau}{B_{0}}k_{\theta n}\dfrac{\omega_{*{\rm i},n}}{\omega_{n}}\delta\phi_{n}\left(\partial_{r}\delta\phi_{{\rm Z,S}}+\partial_{r}\delta\phi_{{\rm Z,B}}\right),\label{eq:DW1}\\
 & \partial_{t}\chi_{{\rm iZ}}\delta\phi_{{\rm Z,S}}=-\dfrac{c}{B_{0}}\alpha_{{\rm i}}k_{\theta n}\rho_{{\rm ti}}^{2}\partial_{r}\left(\delta\phi_{n}\partial_{r}^{2}\delta\phi_{n}^{*}-\delta\phi_{n}^{*}\partial_{r}^{2}\delta\phi_{n}\right),\label{eq:SZF1}\\
 & \chi_{{\rm iZ}}\delta\phi_{{\rm Z,B}}=\dfrac{c}{B_{0}}k_{\theta n}\dfrac{\omega_{*{\rm i},n}}{\omega_{n}^{2}}\left|\partial_{r}\delta\phi_{n}\right|^{2}.\label{eq:BZF1}
\end{eqnarray}
Here, $\mathscr{E}_{{\rm d}}$ is the linear dispersion relation of
DW, $\chi_{{\rm iZ}}$ is the neoclassical inertia enhancement due
to finite drift orbit width effect, $\alpha_{{\rm i}}\equiv1-\omega_{*n,{\rm i}}/\omega_{n}$, and $\rho_{\rm ti}\equiv v_{\rm ti}/\omega_{\rm ci}$.
Eq. (\ref{eq:DW1}) represents the radial envelope modulation of DW
by SZF and BZF; while  Eqs. (\ref{eq:SZF1}) and (\ref{eq:BZF1}) represent
spontaneous  and beat-driven excitation of ZF by DW, respectively. It is noteworthy that, Eqs. (\ref{eq:DW1}), (\ref{eq:SZF1}) and (\ref{eq:BZF1}) have been derived in Ref. \cite{LChenPoP2024}. For the simplicity
of analysis while keeping necessary physics elements, we adopt the
linear dispersion relation of electron DW, which is given by $\omega_{n}\mathscr{E}_{{\rm d}}=\omega-\omega_{*{\rm i},n}-C_{{\rm d}}\omega_{*{\rm i},n}k_{r}^{2}\rho_{{\rm ti}}^{2}$, with $C_{\rm d}$ representing the dispersiveness of DW   \cite{FRomanelliPoFB1993}.
Meanwhile,  $\chi_{{\rm iZ}}\approx1.6q^{2}k_{r{\rm Z}}^{2}\rho_{{\rm ti}}^{2}/\sqrt{\epsilon}$   in the long-wavelength limit of interest \cite{MRosenbluthPRL1998}.
Substituting Eq. (\ref{eq:BZF1}) into DW equation (\ref{eq:DW1}),
 the nonlinear
evolution equations for the radial envelope of DW and SZF are given
by
\begin{eqnarray}
 & \left(\partial_{t}-\gamma_{{\rm L}}-{\rm i}\tau\varOmega(r)-{\rm i}\tau C_{{\rm d}}\partial_{r}^{2}\right)A_{n}={\rm i}\beta A_{n}\delta E_{{\rm Z,S}}\nonumber \\
 & +{\rm i}\dfrac{\sqrt{\epsilon}}{1.6q^{2}}\beta^{2}|A_{n}|^{2}A_{n},\label{eq:DW2}\\
 & \partial_{t}\delta E_{{\rm Z,S}}=-{\rm i}\dfrac{\sqrt{\epsilon}}{1.6q^{2}}\beta\alpha_{{\rm i}}\partial_{r}\left(A_{n}\partial_{r}A_{n}^{*}-A_{n}^{*}\partial_{r}A_{n}\right),\label{eq:SZF2}
\end{eqnarray}
where space and time are normalized by $\rho_{\rm ti}$ and $ 1/\omega_{{\rm *i},n}$, and scalar potential is normalized by $e/T_{\rm i}$.
Here, $\varOmega(r)\equiv\omega_{*{\rm i},n}(r)/\omega_{*{\rm i},n}(0)$
represents the nonuniformity induced by ion diamagnetic well, $\beta\equiv\left(k_{\theta n}\rho_{{\rm ti}}/2\right)\omega_{{\rm ci}}/\omega_{*{\rm i},n}$
is the nonlinear coupling coefficient, $\gamma_{{\rm L}}$ is the linear
growth rate of DW, and $\delta E_{{\rm Z,S}}\equiv-\partial_{r}\delta\phi_{{\rm Z,S}}$
represents the radial electric field of SZF. Note that Eq. (\ref{eq:SZF2}) is the same
as Eq. (4) in Ref. \cite{ZGuoPRL2009}, except for the factor
of $\sqrt{\epsilon}/\left(1.6q^{2}\right)$ decrease in nonlinear
coupling coefficient, connected with inertial enhancement in SZF excitation due to neoclassical
effects in toroidal plasmas. The second term on the right hand side
of Eq. (\ref{eq:DW2}) represents the nonlinear modulation due to
BZF, resulting in a cubic nonlinear term \cite{NChenPoP2024}. In fact, the nonlinear terms
in Eq. (\ref{eq:DW2}) serve as nonlinear potential for DW, given by
$-\beta\delta E_{{\rm Z,S}}-\sqrt{\epsilon}\beta^{2}|A_{n}|^{2}/\left(1.6q^{2}\right)$.
Thus, BZF always contributes as attractive potential well that nonlinearly
traps DW envelopes. Meanwhile, the contribution of SZF depends on its sign/direction,
i.e., $\delta E_{{\rm Z,S}}>0$ for potential well (attractive nonlinearity) and $\delta E_{{\rm Z,S}}<0$
for potential barrier (repulsive nonlinearity). Here, we note that the spatiotemporal spectral features of BZF and SZF are expected to be quite different,  based on their explicit expressions, with the latter weighing more the high-$k_r$ response of DW. When nonlinear trapping induced by  SZF and BZF balance linear dispersiveness of DW, DW soliton can be formed  \cite{ZGuoPRL2009,NChenPoP2024}. The coupled equations (\ref{eq:DW2}) and (\ref{eq:SZF2})
are solved numerically using \textit{pseudospectral} method, with the default
values given by $\beta=10\sqrt{10}$, $\alpha_{\rm i}=\tau=C_{{\rm d}}=1$, and $\sqrt{\epsilon}/\left(1.6q^{2}\right)=0.1$  for typical tokamak parameter regime.
\begin{center}
\begin{figure}
\includegraphics[scale=0.4]{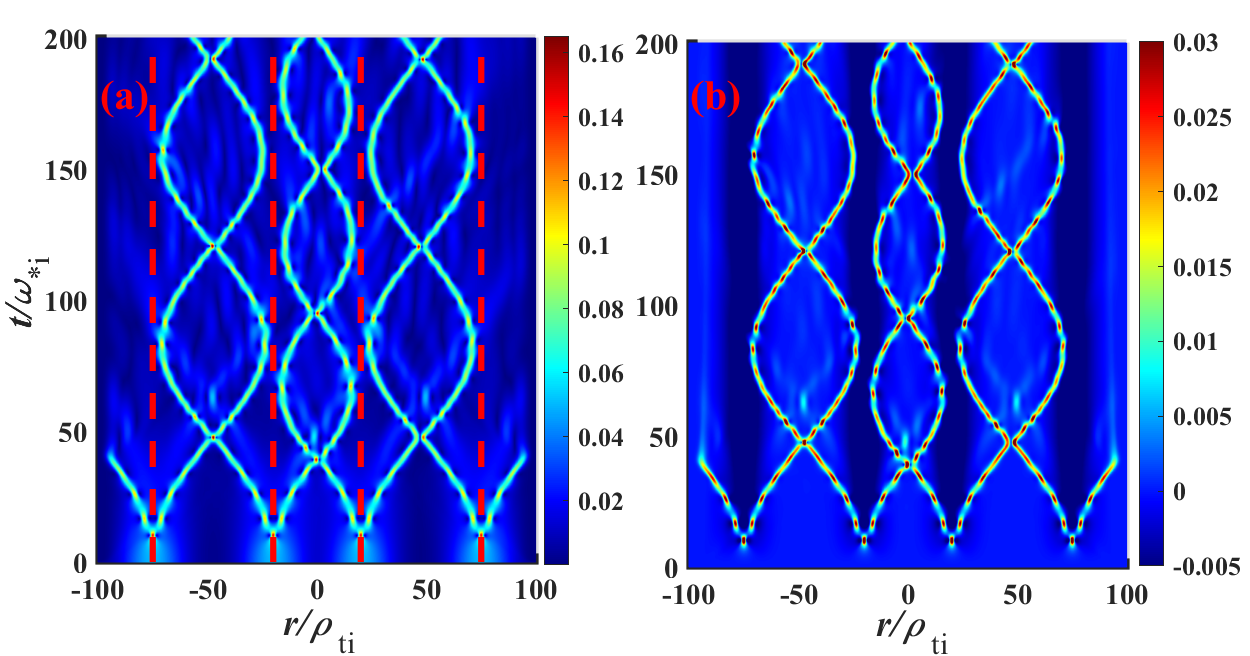}\caption{Spatial-temporal evolution of (a) DW and (b) SZF with regularly spaced
DW pulses with constant $\omega_{*{\rm i}}$, with the red dashed
lines indicating the position of micro-barriers.\label{fig:1}}
\end{figure}
\par\end{center}

We demonstrate first that, in the case with constant $\omega_{*{\rm i}}$,
i.e., $\varOmega(r)=1$, model equations (\ref{eq:DW2}) and (\ref{eq:SZF2})
can produce results similar to the staircase structures observed
in simulations \cite{GPradalierPRL2015,GPradalierNF2017}. Here, regularly spaced DW envelopes with random
phase and noise level SZF are adopted as initial conditions, whose spatial-temporal
evolution are shown in Fig. \ref{fig:1}a and \ref{fig:1}b. The simultaneous
excitation of solitons that enhance turbulence spreading and micro-barriers
localized at corresponding DW peaks that hinder DW solitons from penetrating can
be identified. In fact,
 soliton formation due to ZF excitation was systematically investigated
in Ref. \cite{ZGuoPRL2009,NChenPoP2024}, which demonstrated that
soliton forms as the nonlinear trapping effect induced by ZF excited
by DW balances the dispersiveness. However, in Fig.
\ref{fig:1}b, it is found that SZF contributes  not only to the nonlinear
potential well with $\delta E_{{\rm Z,S}}>0$,
but also to stationary potential barriers with $\delta E_{{\rm Z,S}}<0$
localized around corresponding peaks of the initial DW envelopes, which serve as micro-barriers to barricade
soliton propagation and reflect solitons. Consequently,  staircase-like structure can be observed, with soliton propagation confined within radially spaced micro-barriers denoted by red dashed lines in Fig. \ref{fig:1}a.  Besides, the micro-barriers exhibit long-lived   characteristic,
which is consistent with simulation results \cite{GPradalierPRL2015}. Note  that the reflection
of DW solitons, presented here, can not be due to BZF, which always contributes to attractive nonlinearity as shown above. Thus, BZF contributes to soliton
formation by increasing the depth of nonlinear potential well, and
eventually leads to peaked radial mode structure. Meanwhile, it is clear that SZF can contribute to micro-barrier formation at radial locations where DW intensity peaks and the high-$k_r$ component of the DW spectrum becomes increasingly more important,   consistent with our earlier remarks.

\begin{center}
\begin{figure}
\includegraphics[scale=0.45]{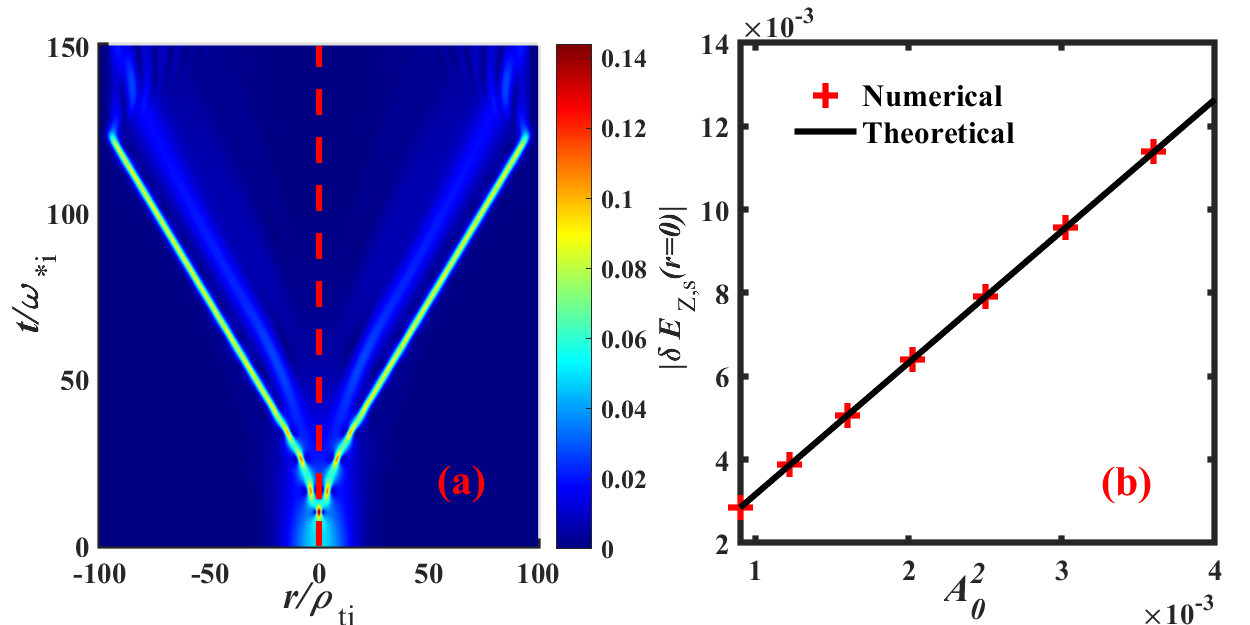}

\caption{(a) Spatial-temporal evolution of DW with single initial pulse, with red dashed line indicating the position of micro-barrier, and
(b) the dependence of asymptotic value of SZF amplitude on the initial
intensity of DW, with the black solid line and red cross indicating
the theoretical and numerical results.\label{fig:2}}
\end{figure}
\par\end{center}

To understand the mechanism of micro-barrier formation, a single DW envelope localized
around the origin is investigated to better delineate underlying physics. It is observed in Fig. \ref{fig:2}a that DW envelope splits
into two counter-propagating solitons, which propagate outward without being reflected, because the micro-barrier, denoted by red dashed line in Fig. \ref{fig:2}a, is localized around peak of initial DW envelope; that is, the origin in this case.
This is a possible reason why the generation of micro-barrier
has not been noted in previous studies on SZF \cite{ZGuoPRL2009}. The simultaneous  generation of potential
wells and barriers by SZF nonlinearity can be analyzed by multiplying $A_{n}^{*}$ with
Eq. (\ref{eq:DW2}), subtracting its complex conjugate, and substituting into  Eq. (\ref{eq:SZF2}). The
nonlinear evolution equation for SZF can be cast as
\begin{eqnarray*}
\partial_{t}\delta E_{{\rm Z,S}} & = & \dfrac{\sqrt{\epsilon}}{1.6q^{2}}\dfrac{\beta\alpha_{{\rm i}}}{\tau C_{{\rm d}}}\partial_{t}\left|A_{n}\right|^{2}.
\end{eqnarray*}
Integrating   the above equation over $t$ and recalling that ZF is noise level
initially, the solution for SZF is given by
\begin{eqnarray}
\delta E_{{\rm Z,S}} & = & \dfrac{\sqrt{\epsilon}}{1.6q^{2}}\dfrac{\beta\alpha_{{\rm i}}}{\tau C_{{\rm d}}}\left[\left|A_{n}\right|^{2}-\left|A_{n}\right|^{2}(t=0)\right],\label{eq:SZF_solution}
\end{eqnarray}
where $\left|A_{n}\right|(t=0)$ is initial envelope of DW. Eq. (\ref{eq:SZF_solution})
demonstrates that SZF excitation is determined by the difference of
DW envelope to its initial value, which, in turn, is a reflection of the available free energy in the considered equilibrium. As the initial envelope splits
due to formation of DW solitons, the amplitude of DW at the origin,
where DW envelope peaks initially, vanishes, and thus, the time asymptotic
value of SZF at $r=0$, i.e, the height  of micro-barriers, is given by
\begin{eqnarray}
\delta E_{{\rm Z,S}}(r=0) & = & -\dfrac{\sqrt{\epsilon}}{1.6q^{2}}\dfrac{\beta\alpha_{{\rm i}}}{\tau C_{{\rm d}}}\left|A_{n}\right|^{2}(t=0,r=0),\label{eq:asymptotic_value}
\end{eqnarray}
which is negative and proportionate to the initial amplitude of DW, accounting for the formation of a repulsive nonlinear term at $r=0$.
By measuring the time asymptotic value of SZF at the origin for different
DW amplitude from numerical studies, the comparison with analytic result given by Eq. (\ref{eq:asymptotic_value}) is shown
in Fig. \ref{fig:2}b, demonstrating excellent agreement. Meanwhile, the radial structure of $\delta E_{{\rm Z,S}}$
is given by $|A_{n}|^{2}$,  which means that the width of micro-barriers is comparable to that of corresponding DW envelope.

Generally speaking, the dual nature of SZF contributing to both attractive and repulsive nonlinear responses in the DW equation is
nearly universal for ZF spontaneously excited by DW from relatively
low residual level. In fact, Eq. (\ref{eq:SZF2}) is essentially a continuity
equation with zero net radial flux, which, after integration over
the whole radial domain, and assuming $A_{n}\rightarrow0$ as $r\rightarrow\infty$,
yields
\begin{eqnarray*}
\partial_{t}\int_{-\infty}^{+\infty}\delta E_{{\rm Z,S}}{\rm d}r & = & 0.
\end{eqnarray*}
Thus, $W\equiv\int_{-\infty}^{+\infty}\delta E_{{\rm Z,S}}{\rm d}r$ is a conserved quantity.
Since SZF is  excited from noise  level initially, we have $W\approx 0$, which means the
excitation of potential well ($\delta E_{{\rm Z,S}}>0$) must be accompanied
by micro-barrier ($\delta E_{{\rm Z,S}}<0$), consistent with
our analysis above. Note that the conservation of $W$ is not affected
by linear properties of DW, i.e., $\mathscr{E}_{{\rm d}}$, but is determined by the structure of
Reynolds stress that nonlinearly excites ZF.  Consequently,
the simultaneous generation of micro-barriers and solitons  is general in nature, and should be observed
in investigations of varieties of DW types, so does its potential
implication, i.e., staircase formation. 

\begin{center}
\begin{figure}
\includegraphics[scale=0.45]{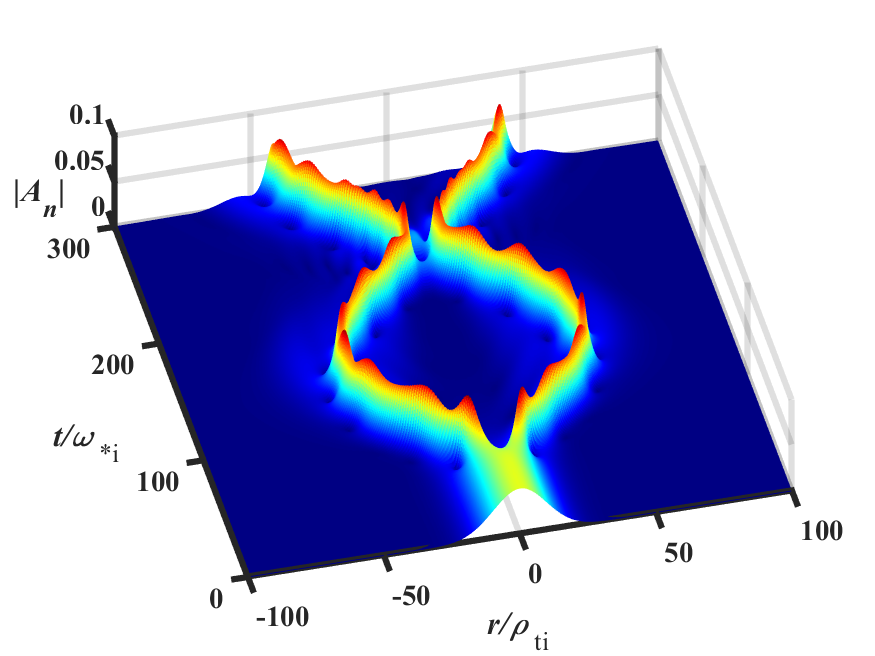}
\caption{Spatial-temporal evolution of a single DW pulse in the nonuniform
diamagnetic well.\label{fig:3}}
\end{figure}
\par\end{center}

In this work, micro-barrier  and staircase structure formation is also investigated
in realistic plasmas with nonuniformity $\varOmega(r)=\exp\left(-r^{2}/L_{{\rm p}}^{2}\right)$
taken into account, where $L_{{\rm p}}$ is the characteristic scale length
of $\omega_{{\rm *i}}\left(r\right)$ variation. It is observed in Fig. \ref{fig:3} that  solitons are reflected back and forth by
the turning points determined by the nonuniform diamagnetic well, and the
nonlinear micro-barrier due to SZF. In this case, a single DW  pulse can itself manifest the formation
of separated DW spreading regions, with the micro-barrier induced by
SZF located in between. 
This is different from the uniform-$\varOmega$ case, where  the propagation of  a DW soliton is limited by the micro-barrier generated by another DW. We however note that the first case investigated, with multiple initial DW pulses with uniform $\varOmega$, is intrinsically nonuniform,  from which multiple DW pulses can exist.

The above-mentioned analyses are based on given initial pulses
of DW  to focus on the physics of micro-barriers formation; while in realistic plasmas, DW is excited from noise level as the system is linearly unstable to DW.  Meanwhile, SZF is excited as DW amplitude is large enough for modulational instability to occur. Without loss of generality, we assume that DW is linearly
excited under spatially uniform $\gamma_{{\rm L}}$, which is artificially
turned off at $t_{{\rm c}}$ to impose ``saturation'', since we have not considered the feedback to $\gamma_{{\rm L}}$ due to ZF induced scattering. Here, $t_{c}$
is the time for ``saturation''. Similar patterns to that shown in Fig.
\ref{fig:1} reappears, which is shown in Fig. \ref{fig:4}. Meanwhile, penetration of micro-barriers can be observed near $r=48$, where the amplitude of soliton is large enough. Recalling that the height and width of micro-barriers are determined by corresponding DW envelopes, present results demonstrate that DW solitons with relatively large amplitude, generated 
near origin in Fig. \ref{fig:4}, can penetrate through micro-barriers generated by lower amplitude DW.

\begin{center}
\begin{figure}
\includegraphics[scale=0.45]{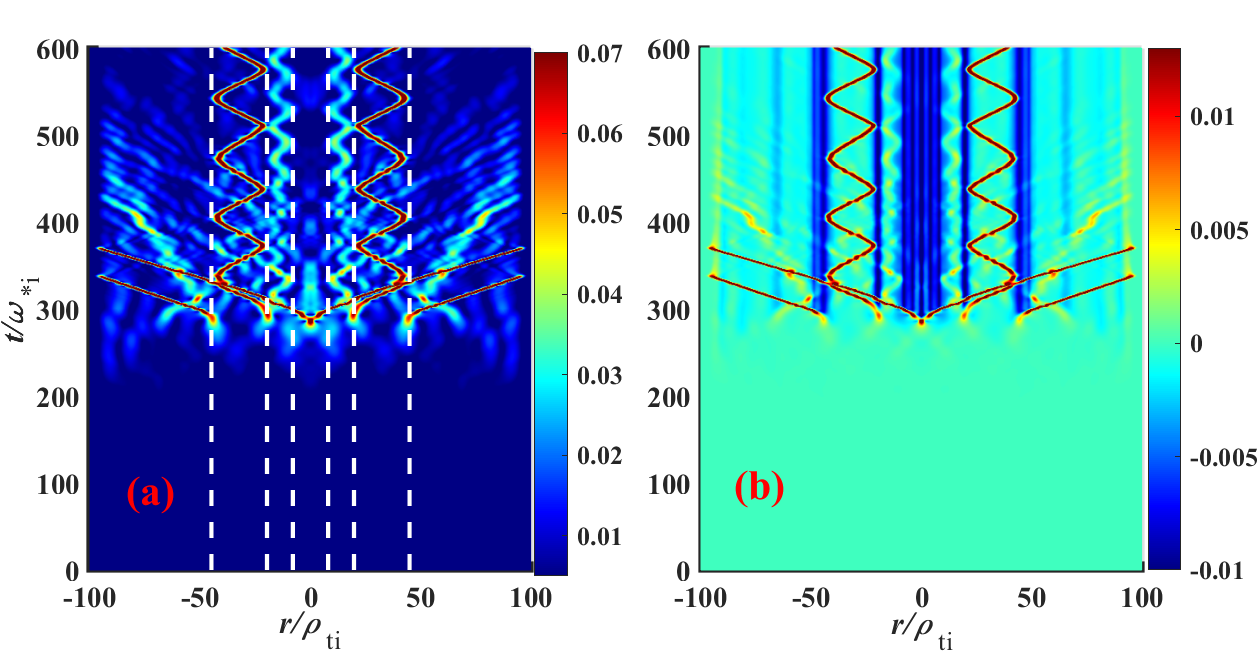}

\caption{Spatial-temporal evolution of (a) DW and (b) SZF with initially noise level
DW suffering finite linear growth rate $\gamma_{{\rm L}}$. White
lines indicate the position of SZF potential barriers. Note here that the initial noise for DW is even symmetric for better visualization, $\gamma_{{\rm L}}=0.025$
and $t_{{\rm c}}=300$. \label{fig:4}}
\end{figure}
\par\end{center}

In summary, we report novel results on simultaneous
exciation of solitons and micro-barriers due to spontaneous excitation
of zonal flow (ZF) by drift wave (DW) using first-principle gyrokinetic
theory, its potential role in explaining $\mathbf{E}\times \mathbf{B}$ staircase formation and relevance to bulk plasma confinement enhancement in magnetically confined fusion devices. DW solitons can be confined within neighboring micro-barriers for initially finite amplitude DW pulses, as well as  for more realistic condition with initially noise level DW with finite linear growth rate. The width and height of micro-barriers are essentially determined by radial mode structure of corresponding DW envelope and, thus, it reflects the available free energy in the considered nonuniform plasma equilibrium. The self-organizing state of confining DW solitons propagation within micro-barriers is beneficial to confinement improvement. Meanwhile, local micro-barriers may be permeable for larger amplitude solitons generated elsewhere.
The dual role of spontaneously excited ZF to act as attractive as well as repulsive nonlinearity in the DW equation is embedded in the structure of Reynolds
stress and zero frequency nature of ZF.  Meanwhile, it is not sensitive to linear dispersion relation of DW, which makes  the present analysis general and its findings should be  applicable to different types of
DW. Geodesic acoustic mode (GAM) can also be excited by Reynolds stress of DW as a finite frequency component of ZF. However, it does not exhibit similar patterns of structure formation due to its finite linear group velocity \cite{ZQiuPPCF2009,NChenPPCF2022}. 

Indeed, this work shed light on understanding the coexistence of solitons
and micro-barriers, and suggests a possible interpretation of the formation of self-organizing
staircase structures, which may improve plasma confinement. However, the characteristics of $\mathbf{E}\times \mathbf{B}$
staircase found in simulations, such as the step size scaling to system size, permeability, and near-marginal
characteristics, are not addressed here. Thus, more detailed exploration via combination of theory and simulation is necessary, which
might be investigated in future publications.

This work was  supported by the Strategic Priority Research Program of Chinese Academy of Sciences under Grant No. XDB0790000,  and  the National Science Foundation of China under Grant Nos. 12275236 and 12261131622.

\end{document}